\documentstyle[aps,prd,%preprint,tighten,
mathrsfs,amsmath,amsfonts,graphicx]{revtex}

\begin{document}

\draft
\preprint{TUM-HEP-405/01}

\title{Neutrino oscillations with three flavors in matter of varying
density}

\author{Tommy Ohlsson\footnote{E-mail address: {\tt
tohlsson@physik.tu-muenchen.de} or {\tt tommy@theophys.kth.se}}}
\address{Institut f{\"u}r Theoretische Physik, Physik Department,
Technische Universit{\"a}t M{\"u}nchen,\\ James-Franck-Stra{\ss}e,
DE-85748 Garching bei M{\"u}nchen, Germany\\ and\\ Division of Mathematical
Physics, Theoretical Physics, Department of Physics, Royal Institute
of Technology,\\ SE-100 44 Stockholm, Sweden}
\author{H{\aa}kan Snellman\footnote{E-mail address: {\tt
snell@theophys.kth.se}}}
\address{Division of Mathematical Physics, Theoretical Physics, Department of
Physics, Royal Institute of Technology,\\ SE-100 44 Stockholm, Sweden}
\date{\today}

\maketitle

\begin{abstract}
In this paper, we discuss the evolution operator and the transition
probabilities expressed as functions of the vacuum mass squared
differences, the vacuum mixing angles, and the matter density
parameter for three flavor neutrino oscillations in matter of varying
density in the plane wave approximation. The applications of this to
neutrino oscillations in a model of the Earth's matter density
profile, step function matter density profiles, constant matter
density profiles, linear matter density profiles, and finally in a
model of the Sun's matter density profile are discussed.
We show that for matter density profiles, which do not fluctuate
too much, the total evolution operator consisting of $n$ operators
can be replaced by one single evolution operator in the semi-classical
approximation.
\end{abstract}


\section{Introduction}
\label{sec:intro}

In previous papers \cite{ohls00,ohls002}, we have given analytic
expressions for the three flavor neutrino oscillation evolution
operator and the transition probabilities in presence of constant
matter densities expressed in the vacuum
mixing matrix elements and the neutrino energies or masses, {\it
i.e.}, incorporating the so-called Mikheyev--Smirnov--Wolfenstein
(MSW) effect \cite{mikh85,wolf78}.
Here we will discuss the application of this to realistic matter density
variations in a ``semi-classical'' approximation based on our previous
results. This allows a simple and efficient calculation of neutrino
oscillations in media of varying densities. We compare this
approximate formula with a numerical simulation in a multi-step model.
We will as before assume that the $CP$ phase $\delta$ is equal to
zero. Thus, the neutrino mixing matrix is real. The semi-classical
approximation for three neutrino flavors, we believe, is a unique part
of our investigation.

Previous work on models for three flavor neutrino oscillations in
matter for constant matter density includes works of Barger {\it et al.}
\cite{barg80},
Kim and Sze \cite{kim87}, and Zaglauer and Schwarzer
\cite{zagl88}. Approximate solutions for three flavor neutrino
oscillations in matter have been presented by Kuo and Pantaleone
\cite{kuo86} and Joshipura and Murthy \cite{josh88}. Approximate
treatments have also been done by Toshev and Petcov \cite{tosh87}.
D'Olivo and Oteo have made contributions by using an approximative
Magnus expansion for the time evolution operator \cite{oliv96}.
Extensive numerical investigations for matter enhanced three neutrino
oscillations have been made by Fogli {\it et al.} \cite{fogl94}.
Studies of neutrino oscillations in Earth has been performed by
several authors \cite{nico88,giun98,liu98,petc98,freu99,moci00}.

Neutrino oscillations for matter with linearly varying density have
been treated by Petcov \cite{petc87} and Lehmann {\it et al.} \cite{lehm00}.
Osland and Wu \cite{osla99} have also solved the case for exponentially
varying density.
Matter enhanced two flavor neutrino oscillations with an arbitrary
monotonic matter density profile have been studied by Balantekin and Beacom
\cite{bala96} using a uniform semi-classical approximation. See also
Fishbane {\it et al.} \cite{fish00} for two flavor neutrino
oscillations in matter of varying density.

\section{The evolution operator in presence of matter}

Let the flavor state basis and mass eigenstate basis be denoted by
${\cal B}_{f} \equiv \{ \vert \nu_\alpha \rangle
\}_{\alpha=e,\mu,\tau}$ and ${\cal B}_{m} \equiv \{ \vert \nu_a \rangle
\}_{a=1}^3$, respectively. Then, the flavor
states $\vert \nu_\alpha \rangle \in {\cal B}_{f}$ can be obtained as a
superpositions of the mass eigenstates $\vert \nu_a \rangle \in {\cal
B}_{m}$, or vice versa. The bases ${\cal B}_f$ and ${\cal
B}_m$ are of course just two different representations of the same
Hilbert space ${\cal H}$.

In the present analysis, we will use the plane wave approximation to
describe neutrino oscillations. In this approximation, a neutrino
flavor state $\vert \nu_{\alpha}\rangle$ is a linear combination of
neutrino mass eigenstates $\vert \nu_{a} \rangle$'s such that \cite{kim93}
\begin{equation}
\vert \nu_\alpha \rangle = \sum_{a=1}^3 U^\ast_{\alpha a} \vert \nu_a
\rangle,
\end{equation}
where $\alpha = e,\mu,\tau$. In what follows, we will
use the short-hand notations $\vert
\alpha \rangle \equiv \vert \nu_\alpha \rangle$ and $\vert a \rangle
\equiv \vert \nu_a \rangle$ for the flavor states and the mass
eigenstates, respectively.

The components of a state $\psi$ in flavor basis and mass
basis, respectively, are related to each other by
\begin{equation}
\psi_f = U \psi_m,
\label{eq:flavor1}
\end{equation}
where
$$
\psi_f \equiv \left( \psi_\alpha \right) \equiv \left( \begin{array}{c} \psi_e
\\ \psi_\mu \\ \psi_\tau
\end{array} \right) \in {\cal B}_f \quad \mbox{and} \quad \psi_m \equiv
\left( \psi_a \right) \equiv \left(
\begin{array}{c} \psi_1 \\ \psi_2 \\ \psi_3 \end{array} \right) \in
{\cal B}_m.
$$

A convenient parameterization for $U = U(\theta_1,\theta_2,\theta_3)$
is given by \cite{groo00}
\begin{equation}
U = \left( \begin{array}{ccc} C_2 C_3 & S_3 C_2 & S_2 \\ - S_3 C_1 -
S_1 S_2 C_3 & C_1 C_3 - S_1 S_2 S_3 & S_1 C_2 \\ S_1 S_3 - S_2
C_1 C_3 & - S_1 C_3 - S_2 S_3 C_1 & C_1 C_2 \end{array} \right),
\end{equation}
where $S_i \equiv \sin \theta_i$ and $C_i \equiv \cos \theta_i$ for $i
= 1,2,3$. This is the standard representation of the neutrino
mixing matrix. The quantities $\theta_i$, where $i = 1,2,3$, are
the vacuum mixing angles. Since we have put the $CP$
phase equal to zero in the neutrino mixing  matrix, this means that
$U^\ast_{\alpha a} = U_{\alpha a}$ for $\alpha = e, \mu, \tau$ and $a = 1,2,3$.

In mass basis, the Hamiltonian ${\mathscr H}$ for the
propagation of the neutrinos in vacuum is diagonal and given by
\begin{equation}
H_m = \left( \begin{array}{ccc} E_{1} & 0 & 0 \\ 0 & E_{2} & 0 \\ 0 &
 0 & E_{3} \end{array} \right),
\end{equation}
where $E_a = \sqrt{m_a^2 + {\bf p}^2}$, $a = 1,2,3$, are the energies
of the neutrino mass eigenstates $\vert a \rangle$, $a = 1,2,3$ with
masses $m_a$, $a = 1,2,3$. We will assume the three-momentum ${\bf p}$
to be the same for all mass eigenstates.

When neutrinos propagate in ordinary matter, there is an additional term in the
Hamiltonian ${\mathscr H}$ coming from the presence of electrons in matter
\cite{wolf78}. This term, the potential term, is diagonal in flavor
basis and is given by
\begin{equation}
V_{f} = A \left( \begin{array}{ccc} 1 & 0 & 0 \\ 0 & 0 & 0 \\ 0 & 0 & 0
\end{array} \right) \equiv A K_{f},
\label{vf}
\end{equation}
where
$$
A \equiv A(r) = \pm \sqrt{2}G_F N_{e}(r) \simeq \pm \frac{1}{\sqrt{2}} G_F
\frac{1}{m_N} \rho(r)
$$
is the matter density parameter and $K_{f}$ is the projector in flavor basis
on the electron neutrinos. Here $G_F$ is the Fermi weak coupling
constant, $N_{e}$ is the electron
density, $m_N$ is the nucleon mass, and $\rho$ is the matter density.
The sign of the matter density parameter depends on weather we deal
with neutrinos~($+$) or antineutrinos~($-$). In mass basis, this piece of
the Hamiltonian is $V_{m}=U^{-1}V_{f}U$, where $U$ is again the
neutrino mixing matrix.

In the case when the neutrinos propagate through matter, as here, the
Hamiltonian
is not diagonal in either the mass basis or the flavor basis, and we have
to calculate the
evolution operator $U_{f}(t)$ or $U_f(L) \equiv e^{-i {\mathscr H}_f L} = U
e^{-i {\mathscr H}_m L} U^{-1}$ if we set $t=L$ ($L$ is the
traveling (propagation) path length of the neutrinos.).

To do so it is convenient to introduce the traceless real symmetric matrix $T$
defined by $T \equiv {\mathscr H}_{m} - ({\rm tr\,} {\mathscr H}_{m}) I/3$.
The trace of the Hamiltonian in mass basis
${\mathscr H}_m \equiv H_m + U^{-1} V_f U$ is
${\rm tr\,} {\mathscr H}_m = E_1 + E_2 + E_3 + A$,
and the matrix $T$ can then be written as
\begin{equation}
T = (T_{ab}) = \left( \begin{array}{ccc} A U_{e1}^2 - \tfrac{1}{3} A +
\tfrac{1}{3} \left( E_{12} + E_{13} \right) & A U_{e1} U_{e2} &A
U_{e1} U_{e3} \\ A U_{e1} U_{e2} & A U_{e2}^2 - \tfrac{1}{3} A +
\tfrac{1}{3} \left( E_{21} + E_{23} \right) & A U_{e2} U_{e3} \\ A
U_{e1} U_{e3} & A U_{e2} U_{e3} & A U_{e3}^2 - \tfrac{1}{3} A +
\tfrac{1}{3} \left( E_{31} + E_{32} \right)
\end{array} \right),
\end{equation}
where $E_{ab} \equiv E_a - E_b$. Of the six antisymmetric quantities
$E_{ab}$, where $a,b=1,2,3$ and $a \neq b$, only two are linearly independent,
since the $E_{ab}$'s fulfill the relations $E_{ba} = - E_{ab}$ and
$E_{12} + E_{23} + E_{31} = 0$.\footnote{Later, we will use the usual
(vacuum) mass squared differences $\Delta m^2_{21}$ and $\Delta
m^2_{32}$, instead of $E_{21}$ and $E_{32}$, which are related to each
other by $\Delta m_{21}^2 = 2 E_\nu E_{21}$ and $\Delta m_{32}^2 \simeq 2 E_\nu
E_{32}$, where $E_\nu$ is the neutrino energy.}
This means that the evolution operator in mass basis can be written as
\cite{ohls00,ohls002}
\begin{equation}
U_m(L) \equiv e^{-i {\mathscr H}_m L} = \phi e^{-i L T} = \phi
\sum_{a=1}^{3} e^{-iL\lambda_{a}}
\frac{1}{3\lambda_{a}^{2}+c_{1}} \left[ (\lambda_{a}^{2}+c_{1})I
+\lambda_{a}T+T^{2} \right],
\label{eq:UmL}
\end{equation}
where $\phi \equiv e^{-i L ({\rm tr \,}{\mathscr H}_{m})I/3}$,
$\lambda_a$, $a=1,2,3$, are the eigenvalues of the matrix $T$, and $I$
is the $3 \times 3$ identity matrix.\footnote{Using Cayley--Hamilton's
theorem, the exponential of a matrix $M$, $e^M = \sum_{n=0}^\infty
\frac{1}{n!} M^n$ (infinite series), can be written as $e^M =
\sum_{n=0}^{N-1} a_n M^n$ (finite series, $N$ is the dimension of
$M$), where $a_n$ ($n = 0,1,\ldots,N-1$) are some coefficients to be
determined. In this case, since $N = 3$ (the dimension of $T$ is
three), this means that we have $e^{-i L T} = a_0 I - i a_1 L T - a_2
L^2 T^2$, {\it i.e.}, there are no higher power terms of $T$ in $e^{-i
L T}$ than that of order two.}
The coefficients $c_0$, $c_1$, and $c_2$ are all real and the
eigenvalues $\lambda_{a}$, $a = 1,2,3$, can be expressed in closed
form in terms of these \cite{ohls00,ohls002}.

The evolution operator for the neutrinos in flavor basis is thus given by
\begin{equation}
U_{f}(L) = e^{-i {\mathscr H}_{f} L} = U e^{-i {\mathscr H}_{m} L}
U^{-1} = \phi \sum_{a=1}^3 e^{-i L \lambda_a}
\frac{1}{3\lambda_a^2+c_1} \left[ (\lambda_a^2 + c_1)I + \lambda_a \tilde{T} +
\tilde{T}^2 \right],
\label{eq:evol}
\end{equation}
where $\tilde{T} \equiv U T U^{-1}$. Equation~(\ref{eq:evol}) is our
final expression for $U_{f}(L)$.

Since ${\mathscr H}_f = U {\mathscr H}_{m} U^{-1}$, it is clear that
$\tilde{T} = {\mathscr H}_{f} - ({\rm tr \,} {\mathscr H}_{f}) I/3 =
{\mathscr H}_{f} - ({\rm tr \,} {\mathscr H}_{m}) I/3$ due to the
invariance of the trace under transformation of $U$. In fact, the
characteristic equation is also invariant under transformation of $U$
and therefore so are the coefficients $c_0$, $c_1$, $c_2$, and the eigenvalues
$\lambda_1$, $\lambda_2$, $\lambda_3$. However, the expression for
${\mathscr H}_{f}$ is much more complicated than that for ${\mathscr H}_{m}$,
which is the reason why we work with ${\mathscr H}_m$ instead of
${\mathscr H}_f$.

The formula~(\ref{eq:evol}) expresses the time (or $L$) evolution
directly in terms of the mass squared differences and the vacuum mixing
angles without introducing any auxiliary matter mixing angles. By dividing the
density variation in small, approximately constant segments, and using this
formula repeatedly in each segment, we can numerically study neutrino
oscillations in matter with varying density. We will use this method as a
standard test for the semi-classical approximation to the evolution
operator that we study below.

\section{The semi-classical approximation}

The formula given in Eq.~(\ref{eq:evol}) is the evolution operator in
flavor basis for
constant matter density. To handle the case of varying matter density,
let us divide the distance $L$ from the source to the detector into
$N$ equidistant parts and put the index $k$ on the eigenvalues
$\lambda_{a}^{k}$, where $a = 1,2,3$ denote the three mass
eigenstates. For any matter density profile $\rho(r)$ we first make
this profile discrete and introduce $\rho_{k}$ as the matter density
in the interval $r_{k-1} \leq r \leq r_{k}$, where $k$ varies from $1$
to $N$ with $r_{0}=0$ and $r_{N}=L$. The length of each segment is then
$\Delta r_{k}= r_{k}-r_{k-1}=L/N$. The evolution operator in mass
basis from $0$ to $L$ can then be written as the ordered product
\begin{equation}
U_m(L) = U_m(r_{N}-r_{N-1}) U_m(r_{N-1}-r_{N-2}) \ldots
U_m(r_{2}-r_{1}) U_m(r_{1}-r_{0})
\label{evol2}.
\end{equation}
Note that the order of the $U_m(r_k - r_{k-1})$'s are important, since
these operators do not in general commute.

When $N$ is large, each step is small and the exponent in the
evolution operator $U_m(r_{k}-r_{k-1}) = U_m(\Delta r_{k})$ is small. We
can then approximate this operator with
\begin{equation}
U_m(\Delta r_{k}) \simeq  e^{-i\Delta r_{k} {\mathscr H}_{m}^{k}},
\label{evol3}
\end{equation}
where ${\mathscr H}_{m}^{k} \equiv H_{m} + A_{k} K_{m}$ and $A_{k}
\propto \rho_{k}$. Inserting this into Eq.~(\ref{evol2}) gives
\begin{equation}
U_m(L) \simeq e^{-i \Delta r_{N} {\mathscr H}_{m}^{N}} e^{-i \Delta
r_{N-1} {\mathscr H}_{m}^{N-1}} \ldots e^{-i \Delta r_{2}{\mathscr
H}_{m}^{2}} e^{-i \Delta r_{1}{\mathscr H}_{m}^{1}}.
\label{evol4}
\end{equation}
Since the ${\mathscr H}_{m}^{k}$'s do not commute, the higher
order terms have to be calculated with the time-ordering (here rather
$r$-ordering) operator.
However, here we will at first be satisfied with the lowest order result,
which we call the {\it semi-classical approximation}. In this approximation, we
retain only the terms proportional to $\Delta r_{k} = L/N$, and thus,
neglect the noncommutativity of the ${\mathscr H}_{m}^{k}$'s for
different $k$'s. We can thus write
\begin{equation}
U_m(L) \simeq e^{-i \sum_{k = 1}^{N} \frac{L}{N} {\mathscr H}_{m}^{k}}.
\label{evol5}
\end{equation}
In the limit $N\rightarrow \infty$, this gives the integral formula
\begin{equation}
U_m(L) = e^{-i \int_{0}^{L} {\mathscr H}_m(r) \, dr} = \phi(L) e^{-i
\int_{0}^{L} T(r) \, dr},
\label{evol6}
\end{equation}
where $T(r)$ is the traceless part of the Hamiltonian corresponding
to the electron density at position $r$ between $0$ and $L$ and
$\phi(L)$ is the phase factor coming from the trace.

For further discussion it is often convenient to retain the original
form of the Hamiltonian and to use
${\mathscr H}_{m}=H_{m}+AU^{-1}K_{f}U$ rather than $T$.
Thus, when $A=A(r)$, we obtain
\begin{equation}
\int_{0}^{L}{\mathscr H}_{m}(r) \, dr = L(H_{m}+ \bar A(L) K_{m}),
\label{expo1}
\end{equation}
where
$$
{\bar A(L)} \equiv \frac{1}{L}\int_{0}^{L}A(r) \, dr
$$
is the average matter density along the baseline $L$ and
$$
K_{m} \equiv U^{-1}K_{f}U,
$$
which means that the evolution operator in mass basis can be written
as
\begin{equation}
U_m(L) = e^{-i L \left(H_m + {\bar A}(L) K_m\right)} \equiv {\bar
\phi} e^{-i L {\bar T}},
\end{equation}
where $\bar{\phi} \equiv e^{-i L ({\rm tr \,} \bar{{\mathscr
H}}_m)/3}$, $\bar{T} \equiv \bar{{\mathscr H}}_m - ({\rm tr \,}
\bar{{\mathscr H}}_m) I/3$, and $\bar{{\mathscr H}}_m \equiv H_m +
\bar{A}(L) K_m$.

For this case we can thus use the previous expression for
$T ={\mathscr H}_{m} - ({\rm tr \,}{\mathscr H}_{m}) I/3 $ by simply
replacing $A$ with $\bar A(r)$ and then pass to flavor basis by using
the $U$ transformation, {\it i.e.}, ${\tilde T}=UTU^{-1}$.

Thus, for any $L$ we can use the spectral decomposition theorem and we
find that
\begin{equation}
U_f(L) =\phi(L)\sum_{a=1}^{3}e^{-iL\lambda_{a}(L)}P_{a}(L),
\label{spec1}
\end{equation}
where $\lambda_{a}(L)$ is the $a$th eigenvalue of $T(L)$ (or ${\tilde
T(L)}$) and
\begin{equation}
P_{a}(L)= \frac{1}{3\lambda_{a}^{2}(L)+c_{1}(L)}
\left[(\lambda_{a}^{2}(L)+c_{1}(L))I +
\lambda_{a}(L)\tilde{T}(L)+\tilde{T}^{2}(L) \right]
\label{proj1}
\end{equation}
is the projection operator.
Everything here is of course $L$-dependent, since
the operator is $L$-dependent and therefore also the eigenvalues. The
phase factor is $\phi(L)=e^{-iL ({\rm tr \,}{\mathscr H}_{f}(L))/3}
= e^{-iL ({\rm tr \,}{\mathscr H}_{m}(L))/3}$.

In the case of a linear matter density of the form $A(r)=A+Br$, we obtain
$\bar A(r) =A+Br/2$.
Similarly, in the case of a step function like matter density, relevant to the
matter distribution of the Earth, we have
$$
A(r) = \left\{ \begin{array}{ll} A_1, & 0 \leq r \leq L_1, \\ A_2, &
L_1 \leq r \leq L_1 + L_2, \\ A_1, & L_1 + L_2 \leq r \leq 2 L_1 +
L_2, \end{array} \right., \quad \mbox{where $2 L_1 + L_2 \equiv L$,}
$$
which leads to
\begin{equation}
\bar{A}(r) = \left\{ \begin{array}{ll} A_1, & 0 \leq r
\leq L_1,\\ A_2 \left( 1 - \frac{L_1}{r} \right) + A_1 \frac{L_1}{r},
& L_1 \leq r \leq L_1 + L_2,\\ A_1 \left( 1 - \frac{L_2}{r} \right) +
A_2 \frac{L_2}{r}, & L_1 + L_2 \leq r \leq 2 L_1 + L_2, \end{array} \right..
\end{equation}
Finally, in the case of an exponentially decreasing matter density $A(r) = A
e^{-r/r_{0}}$, where $A$ and $r_{0}$ are parameters relevant to
the matter distribution of the Sun, we obtain
\begin{equation}
\bar{A}(r) = A \frac{r_{0}}{r} \left(1 - e^{-r/r_{0}}\right).
\end{equation}

We can see here that in the semi-classical approximation the influence of
the density $\bar A$ decays as $1/L$ with distance $L$ from the matter. The
evolution should therefore be continued with the vacuum evolution operator
as soon as the neutrinos leave the matter region.

\section{Digression on the semi-classical approximation}

Let us consider the semi-classical (s.c.) evolution operator further. It can
be written as
\begin{equation}
U(r)_m^{\rm s.c.} = e^{-i{\mathscr H}^{(1)}_{m}(r)},
\label{sc10}
\end{equation}
where
$$
{\mathscr H}^{(1)}_{m}(r) \equiv H_{m}r + A^{(1)}(r) K_{m}.
\label{scham}
$$
Here $A^{(1)}(r) \equiv \int_0^r A(s) \, ds$.
Now, the equation of motion for the full evolution operator $U_m$ is
\begin{equation}
i\frac{d}{dr}U_m(r) = {\mathscr H}_{m}(r) U_m(r).
\label{em}
\end{equation}
This can be integrated to give the equation
\begin{equation}
U_m(r) = 1 - i \int_{0}^{r} {\mathscr H}_{m}(s) U_m(s) \, ds.
\label{em2}
\end{equation}
Upon differentiating the semi-classical evolution operator above,
we see that, although
\begin{equation}
\frac{d}{dr}{\mathscr H}^{(1)}_m(r) = {\mathscr H}_{m}(r) = H_{m} + A(r) K_{m},
\label{derivativeh}
\end{equation}
we can equate $i \frac{d}{dr} U_m^{\rm s.c.}(r)$ with ${\mathscr
H}_{m}(r) U_m^{\rm s.c.}(r)$ only when the commutator $\left[{\mathscr
H}^{(1)}_m(r),{\mathscr H}_{m}(r)\right]$ can be neglected. This commutator
can be calculated to be
\begin{equation}
\left[{\mathscr H}^{(1)}_m(r),{\mathscr H}_{m}(r)\right] = \int_{0}^{r}
s \frac{dA}{ds}(s) \, ds \left[H_{m},K_{m}\right].
\label{commutator}
\end{equation}
Thus, when $\int_{0}^{r} s\frac{dA}{ds}(s) \, ds = \frac{r^{2}}{2}
\frac{dA}{ds}(\xi)$ for $0\leq \xi\leq r$ is small, the semi-classical
approximation to the evolution operator is a good approximation to the
full evolution operator. For constant matter density this is of course
true. For linear matter density $A(r)=A+Br$ the coefficient $B$ should
be small, {\it i.e.}, $Br \ll A$ or at least $Br < A$.

Equation~(\ref{em2}) can be solved in a systematic
way by iteration, leading to
\begin{equation}
U_m(r) = 1 - i\int_{0}^{r} {\mathscr H}_{m}(s) \, ds
+ (-i)^{2} \int_{0}^{r} {\mathscr H}_{m}(s) \int_{0}^{s}{\mathscr
H}_{m}(s') \, ds' \, ds + \ldots.
\label{em3}
\end{equation}
The Hamiltonian ${\mathscr H}_{m}(r)$ is the one given in
Eq.~(\ref{derivativeh}). The result to second order is then given by
\begin{equation}
U_m(r) \simeq 1 - i r \left( H_{m} + {\bar A}(r) K_{m} \right) +
(-i)^{2} \frac{r^{2}}{2} \left( H_{m} + {\bar A}(r) K_{m} \right)^{2} +
(-i)^{2} \frac{r^{2}}{2} \left( {\bar{\bar A}}(r) - {\bar A}(r) \right)
\left[H_{m},K_{m}\right],
\label{em4}
\end{equation}
where
$$
{\bar A}(r) \equiv A^{(1)}(r)/r, \qquad
A^{(1)}(r) \equiv \int_{0}^{r}A(s) \, ds, \qquad
{\bar {\bar A}}(r) \equiv 2A^{(2)}(r)/r^{2}, \qquad
A^{(2)}(r) \equiv \int_{0}^{r}A^{(1)}(s) \, ds.
$$
By inspection we see that the expression in Eq.~(\ref{em4}) deviates from
an expansion of the
semi-classical approximation by the terms proportional to the
commutator $\left[H_{m},K_{m}\right]$ and higher order terms in $H_{m}$ and
$K_{m}$. In
fact, the commutator between the Hamiltonian at different points $s$
and $s'$ is
\begin{equation}
\left[{\mathscr H}_{m}(s),{\mathscr H}_{m}(s')\right] = (A(s)-A(s'))
\left[H_{m},K_{m}\right] = \left.\frac{dA}{ds}\right\vert_{s = s'}(s -
s') [H_m,K_m] + \ldots,
\label{comm1}
\end{equation}
which vanishes only for $A(s)=A(s')$. This is in general true only
for constant matter densities, $A(s)=A$. When $\frac{dA}{ds}$ is
large, the contribution of the commutator cannot be neglected.

We can therefore sum the semi-classical approximation terms and write
the solution as
\begin{equation}
U_{m}(r) = U_m^{\rm s.c.}(r) + a_{1} \left[H_{m},K_{m}\right] + \ldots,
\label{solution10}
\end{equation}
where
$$
a_{1} = (-i)^{2}\frac{r^{2}}{2} \left({\bar{\bar A}}(r) - {\bar A}(r)\right)
$$
and the dots represent higher order terms that vanish when the
commutator $\left[H_{m},K_{m}\right]$ is neglected. When $\left\vert
{\bar{\bar A}} - {\bar A} \right\vert$
is small, the correction terms are small.

\section{Probability amplitudes and transition probabilities}
\label{sec:prob}

In the previous sections, we have calculated the evolution
operator in the semi-classical approximation. Below we will study the
corresponding probability amplitudes and transition probabilities.

The probability amplitude $A_{\alpha\beta}$ for $\nu_\alpha \to
\nu_\beta$ transition is simply defined as the $(\beta,\alpha)$-matrix
element of the evolution operator in flavor basis, {\it i.e.},
\begin{equation}
A_{\alpha\beta} \equiv \langle \beta \vert U_f(L) \vert \alpha
\rangle, \quad \alpha,\beta=e,\mu,\tau.
\label{eq:ampl}
\end{equation}
We now consider transition probabilities for neutrino oscillations
in the semi-classical approximation given by Eq.~(\ref{spec1}).
Inserting Eq.~(\ref{spec1}) into Eq.~(\ref{eq:ampl}) gives
\begin{equation}
A_{\alpha\beta} = \phi(L) \sum_{a=1}^3 e^{-i L
\lambda_a(L)}P_{a}(L)_{\beta\alpha}
\label{eq:ampl2}
\end{equation}
where
\begin{equation}
P_{a}(L)_{\beta\alpha}=\frac{(\lambda_a^2 + c_1)
\delta_{\beta\alpha} + \lambda_a \tilde{T}_{\beta\alpha} +
(\tilde{T}^2)_{\beta\alpha}}{3\lambda_a^2+c_1}
\label{ampl1}
\end{equation}
is the matrix element of the projector $P_{a}(L)$. Here $\delta_{\alpha\beta}$
is Kronecker's delta. Note that $\tilde{T}_{\alpha\beta} =
\tilde{T}_{\beta\alpha}$ and $(\tilde{T}^2)_{\alpha\beta} =
(\tilde{T}^2)_{\beta\alpha}$.
The transition probability $P_{\alpha\beta}$ for $\nu_\alpha \to
\nu_\beta$ transition is defined as the absolute value squared of the
probability amplitude $A_{\alpha\beta}$.
Hence, the transition probabilities in matter are given by the formulas
\begin{equation}
P_{\alpha\beta} = \vert A_{\alpha\beta} \vert^2 = \delta_{\alpha\beta}
- 4 \; \underset{a < b}{\sum_{a=1}^3
\sum_{b=1}^3} P_{a}(L)_{\beta\alpha}P_{b}(L)_{\beta\alpha}
\sin^{2}\tilde{x}_{ab}, \quad \alpha,\beta = e,\mu,\tau,
\label{eq:Pab_2}
\end{equation}
where $\tilde{x}_{ab} \equiv (\lambda_a(L) - \lambda_b(L))L/2$.

\section{Applications and Discussion}

The main results of our analysis are given by the evolution
operator for the neutrinos when passing through matter with varying matter
density in the ``semi-classical approximation'', Eq.~(\ref{spec1}),
and the corresponding expressions for the transition amplitudes in
Eq.~(\ref{eq:ampl2}) and the transition probabilities in
Eq.~(\ref{eq:Pab_2}), all expressed as finite sums of simple functions
in the matrix elements of ${\mathscr H}_{f}$ (or ${\mathscr H}_{m}$)
and integrals involving $A(r)$, the varying matter density.

As applications, we have calculated the transition probability
$P_{\mu e}$ for neutrino oscillations for different matter density
profiles. Our calculations compare two different cases:
\begin{enumerate}
\item An ``exact'' numerical evolution operator method based on the
product of $N$ evolution operators using
\begin{equation}
U_f(L) = U_f(r_{N}-r_{N-1}) U_f(r_{N-1}-r_{N-2}) \ldots
U_f(r_{2}-r_{1}) U_f(r_{1}-r_{0}),
\label{evol12}
\end{equation}
with the formula~(\ref{eq:evol}) used for each step with the appropriate
matter density; and
\item The semi-classical approximation method based on one single
evolution operator using
\begin{equation}
U_f(L) = \bar{\phi} U e^{-i L \bar{T}} U^{-1}.
\end{equation}
\end{enumerate}

In all our examples discussed below, we have used the Earth center
crossing neutrino traveling path length, except for the last example
in which we discuss the Sun.

Let $R_\oplus \simeq 6371$ km be the radius of
the Earth and $r_\oplus \simeq 3486$ km be the radius of the core with this
approximation and in a numerical simulation based on Eq.~(\ref{evol2})
with a step length of $L/N = 2R_\oplus/100 = 127.42$ km. The thickness of the
mantle is then $R_\oplus-r_\oplus \simeq 2885$ km with the matter density
parameter $A_{1} = A_{\rm mantle} \simeq 1.70 \cdot 10^{-13}$ eV ($\rho_1 =
\rho_{\rm mantle} \simeq 4.5$ ${\rm g/cm^3}$), whereas the matter
density parameter of the core is $A_{2} = A_{\rm core} \simeq 4.35
\cdot 10^{-13}$ eV ($\rho_2 = \rho_{\rm core} \simeq 11.5$ ${\rm g/cm^3}$).

Neutrinos traversing the Earth towards a detector close
to the surface of the Earth, pass through the matter of varying density
densities $A(r)$ where the distances
$L_{i}$, $i=1,2$, are functions of the nadir angle $h$, where $h \equiv
\pi-\theta_{z}$; $\theta_{z}$ being the zenith angle. As $h$ varies
from $0$ to $\pi/2$, the cord $L = L(h)$ of the neutrino passage through
Earth becomes shorter and shorter. At an angle larger than $h_{0} =
\arcsin (r_\oplus/R_\oplus) \simeq 33.17^{\circ}$, the distance
$L_{2}=0$, and the neutrinos no longer traverse the core.

The mass squared differences ($\Delta M^{2} \equiv \Delta m_{32}^2$
and $\Delta m^{2} \equiv \Delta m_{21}^2$) and
the vacuum mixing angles ($\theta_{1},\theta_{2},\theta_{3}$) used here
are chosen to correspond to those obtained from analyses of various neutrino
oscillation data. We have taken
$$
\Delta M^{2} = 3.2 \cdot 10^{-3} {\rm eV}^{2}, \quad
\Delta m^{2} = 0, \quad
\theta_{1} = 45^{\circ}, \quad
\theta_{2} = 5^{\circ}, \quad
\theta_{3} = 45^{\circ}.
$$
The values of $\Delta M^{2}$ and $\theta_{1}$ are governed by
atmospheric neutrino data \cite{scho99} and the values of $\Delta m^{2}$ and
$\theta_{3}$ (LMA) by solar neutrino data \cite{bahc98}, where LMA
stands for large mixing angle (matter) solution. The value of $\theta_{2}$ is
below the CHOOZ upper bound, which is $\sin^2 2\theta_2 = 0.10$ \cite{apol98}.
These choices are the most optimistic ones for obtaining any effects
in long baseline (LBL) experiments from the sub-leading $\Delta m^{2}$ scale
\cite{barg00}. We should mention though, that these data are taken from
two neutrino flavor model analyses.

\begin{figure}
\begin{center}
\includegraphics[height=8cm,angle=-90]{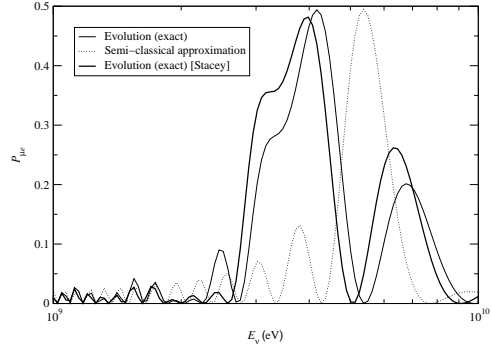}
\end{center}
\caption{The transition probability $P_{\mu e}$ as a function of the
neutrino energy $E_\nu$ for the mantle-core-mantle step function approximation
of the Earth's matter density profile. The mean matter density of the
mantle and the core were chosen to be $\rho_{\rm mantle} = 4.5 \,
{\rm g/cm^3}$ ($A_{\rm mantle} \simeq 1.7 \cdot 10^{-13} \, {\rm eV}$,
$L_{\rm mantle} = 2885 \, {\rm km}$) and $\rho_{\rm core} = 11.5 \,
{\rm g/cm^3}$ ($A_{\rm core} \simeq 4.4 \cdot 10^{-13} \, {\rm eV}$,
$L_{\rm core} = 6972 \, {\rm km}$), respectively. Parameter values: $h
= 0$, $\theta_1 = 45^\circ$, $\theta_2 = 5^\circ$, $\theta_3 =
45^\circ$, $\Delta m^2 = 0$, and $\Delta M^2 = 3.2 \cdot 10^{-3} \,
{\rm eV}^2$.}
\label{fig1}
\end{figure}

As a first example, we have investigated the Earth's matter density
profile using the published Stacey model for the Earth's matter
density profile \cite{stac77}. The resulting curves of the (exact)
numerical evolution operator method for a mantle-core-mantle step function
approximation of the Earth's matter density profile, the
semi-classical approximation method, and for reference the (exact) numerical
evolution operator method of the Earth's matter density profile are
shown in Fig.~\ref{fig1}.
It has earlier been found by Freund and Ohlsson \cite{freu99} that a
mantle-core-mantle step function approximation of the Earth's matter
density profile is a good approximation (even in a three neutrino scenario).
The numerical evolution operator method results were carried out using
$N = 100$, {\it i.e.}, they consist of a product of 100 evolutions
with different constant matter densities for each evolution step, whereas the
semi-classical approximation method result was obtained with just one single
evolution with the average matter density of the Earth's matter
density profile $\bar{A}_\oplus$ ($\bar{\rho}_\oplus \simeq 7.8 \,
{\rm g/cm^3}$, which is also the reason why
the semi-classical approximation curve only has got one resonance peak
at $E_\nu = \bar{E}_{\nu,\oplus} \simeq 5.4 \cdot 10^9 \, {\rm eV}$. This
peak of course lies inbetween the both resonance peaks (ideally at $E_\nu =
E_{\nu, {\rm core}} \simeq 3.7 \cdot 10^9 \, {\rm eV}$ and $E_\nu =
E_{\nu, {\rm mantle}} \simeq 9.4 \cdot 10^9 \, {\rm eV}$) of the exact
numerical evolution calculation, since $A_{\rm mantle} \leq
\bar{A}_\oplus \leq A_{\rm core}$.

\begin{figure}
\begin{center}
\includegraphics[height=8cm,angle=-90]{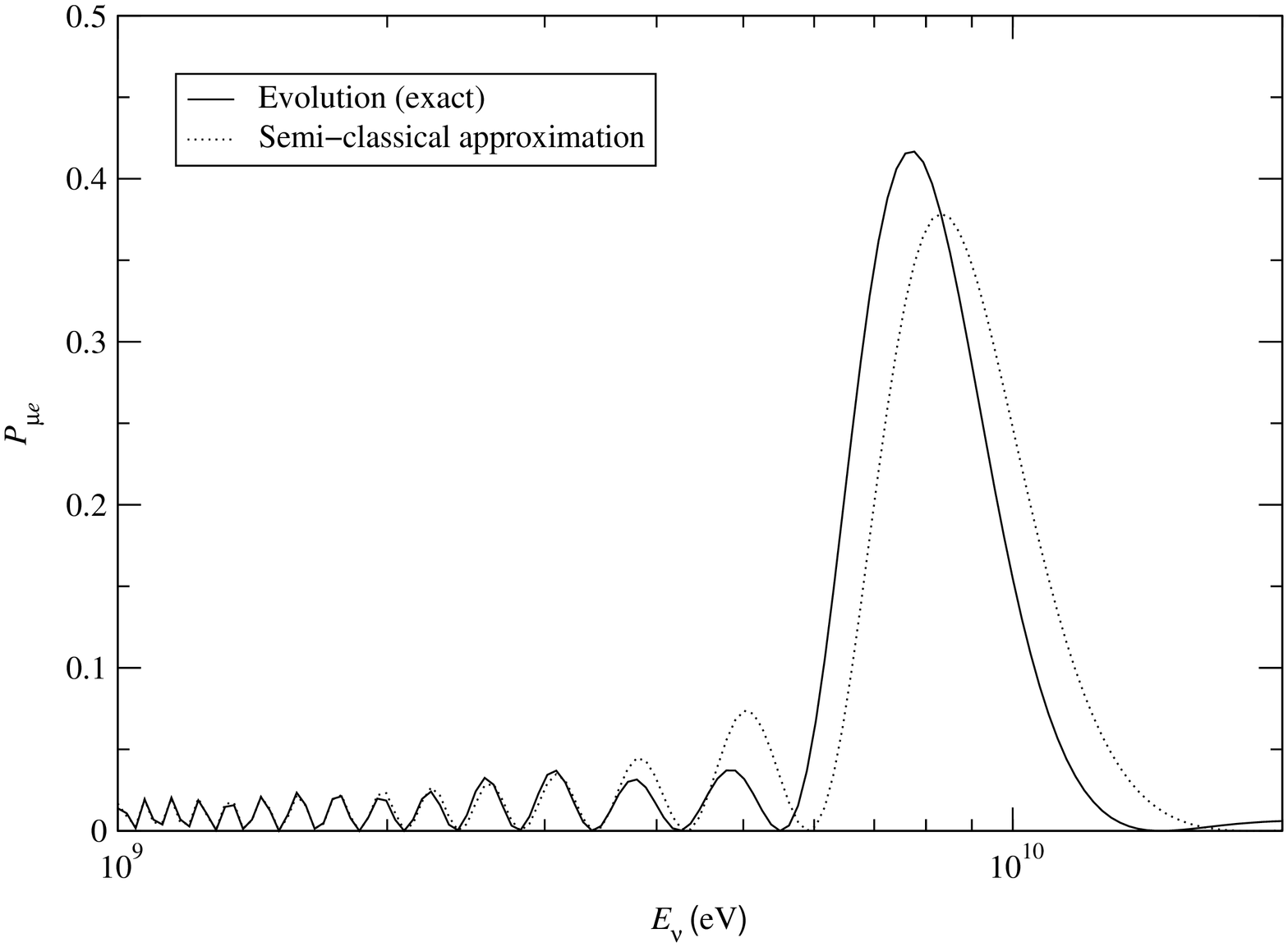}
\end{center}
\caption{The transition probability $P_{\mu e}$ as a function of the
neutrino energy $E_\nu$ for a step function matter density profile with $\rho_1
= 4.5 \, {\rm g/cm^3}$ ($A_1 \simeq 1.7 \cdot 10^{-13} \, {\rm eV}$,
$L_1 = 2885 \, {\rm km}$) and $\rho_2 = 5.5 \, {\rm g/cm^3}$ ($A_2
\simeq 2.1 \cdot 10^{-13} \, {\rm eV}$, $L_2 = 6972 \, {\rm
km}$). Parameter
values: $h = 0$, $\theta_1 = 45^\circ$, $\theta_2 = 5^\circ$, $\theta_3 =
45^\circ$, $\Delta m^2 = 0$, and $\Delta M^2 = 3.2 \cdot 10^{-3} \,
{\rm eV}^2$.}
\label{fig2}
\end{figure}

\begin{figure}
\begin{center}
\includegraphics[height=8cm,angle=-90]{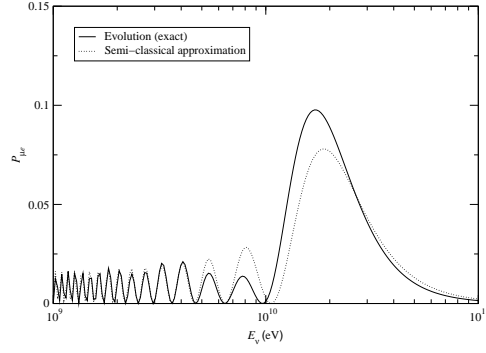}
\end{center}
\caption{The transition probability $P_{\mu e}$ as a function of the
neutrino energy $E_\nu$ for a step function matter density profile with $\rho_1
= 1 \, {\rm g/cm^3}$ ($A_1 \simeq 3.8 \cdot 10^{-14} \, {\rm eV}$,
$L_1 = 2885 \, {\rm km}$) and $\rho_2 = 2 \, {\rm g/cm^3}$ ($A_2
\simeq 7.6 \cdot 10^{-14} \, {\rm eV}$, $L_2 = 6972 \, {\rm
km}$). Parameter
values: $h = 0$, $\theta_1 = 45^\circ$, $\theta_2 = 5^\circ$, $\theta_3 =
45^\circ$, $\Delta m^2 = 0$, and $\Delta M^2 = 3.2 \cdot 10^{-3} \,
{\rm eV}^2$.}
\label{fig3}
\end{figure}

In Figs.~\ref{fig2} and \ref{fig3}, we have used two different step
function matter density profiles.

There appear no double peaks in these figures even though the step
function matter density profiles consist of two different
$A_k$'s. Furthermore, in Figs.~\ref{fig2} and \ref{fig3}, the
differences between the values of the $A_1$'s and $A_2$'s are the
same. The step
function matter density profile in Fig.~\ref{fig2} could simulate the
Earth's matter density profile if the Earth has a core, which is much
less dense than has been found by geophysics. Note that the absolute
error between the two curves in Fig.~\ref{fig2} is larger than in
Fig.~\ref{fig3}, whereas the relative error of the curves in
Fig.~\ref{fig3} is larger than in Fig.~\ref{fig2}.

\begin{figure}
\begin{center}
\includegraphics[height=8cm,angle=-90]{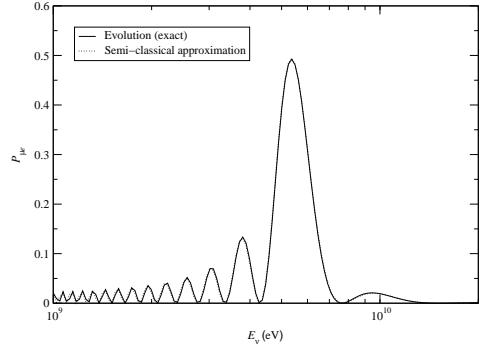}
\end{center}
\caption{The transition probability $P_{\mu e}$ as a function of the
neutrino energy $E_\nu$ for a constant matter density profile with $\rho
= 7.8 \, {\rm g/cm^3}$ ($A \simeq 3.0 \cdot 10^{-13} \, {\rm eV}$, $L
= 12742 \, {\rm km}$). Parameter values: $h = 0$, $\theta_1 =
45^\circ$, $\theta_2 = 5^\circ$, $\theta_3 = 45^\circ$, $\Delta m^2 =
0$, and $\Delta M^2 = 3.2 \cdot 10^{-3} \, {\rm eV}^2$.}
\label{fig4}
\end{figure}

\begin{figure}
\begin{center}
\includegraphics[height=8cm,angle=-90]{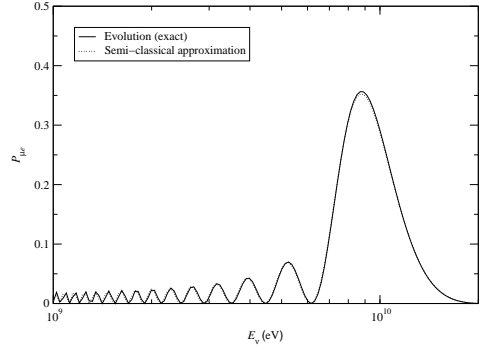}
\end{center}
\caption{The transition probability $P_{\mu e}$ as a function of the
neutrino energy $E_\nu$ for a constant matter density profile with $\rho
= 4.5 \, {\rm g/cm^3}$ ($A \simeq 1.7 \cdot 10^{-13} \, {\rm eV}$, $L
= 12742 \, {\rm km}$). Parameter values: $h = 0$, $\theta_1 =
45^\circ$, $\theta_2 = 5^\circ$, $\theta_3 = 45^\circ$, $\Delta m^2 =
0$, and $\Delta M^2 = 3.2 \cdot 10^{-3} \, {\rm eV}^2$.}
\label{fig5}
\end{figure}

Next, in Figs.~\ref{fig4} and \ref{fig5}, we have studied constant
matter density profiles.

The constant matter density used in Fig.~\ref{fig4} is the average
density of the Earth. The curves in this figure could be compared with
the dotted curve in Fig.~\ref{fig1}. %good approx.
In Fig.~\ref{fig5}, the constant matter density was chosen to be equal
to the average density in the mantle of the Earth. In both these
figures, the semi-classical approximation method gives an excellent agreement
with the exact numerical evolution operator method as it should, since
in the case of constant matter density $\bar{A} = A = {\rm const.}$,
which means that the two methods are equivalent. Thus, the very small
deviations seen in the figures are only due to numerics.

\begin{figure}
\begin{center}
\includegraphics[height=8cm,angle=-90]{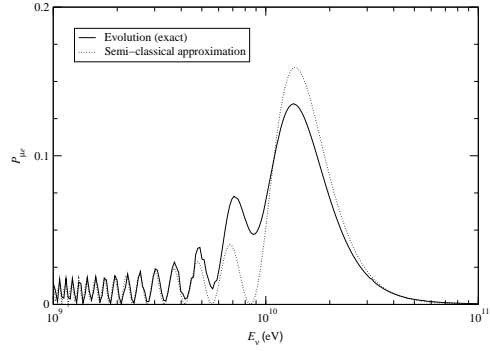}
\end{center}
\caption{The transition probability $P_{\mu e}$ as a function of the
neutrino energy $E_\nu$ for a linear matter density profile with $A =
0$ and $BL \simeq 3.8 \cdot 10^{-13} \, {\rm eV}$ (corresponding to
$\rho = 5 \, {\rm g/cm^3}$ and $L = 12742 \, {\rm km}$ if $A =
\frac{BL}{2}$). Parameter values:
$h = 0$, $\theta_1 = 45^\circ$, $\theta_2 = 5^\circ$, $\theta_3 = 45^\circ$,
$\Delta m^2 = 0$, and $\Delta M^2 = 3.2 \cdot 10^{-3} \, {\rm eV}^2$.}
\label{fig6}
\end{figure}

\begin{figure}
\begin{center}
\includegraphics[height=8cm,angle=-90]{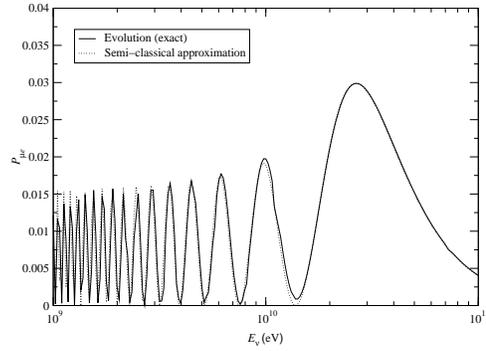}
\end{center}
\caption{The transition probability $P_{\mu e}$ as a function of the
neutrino energy $E_\nu$ for a linear matter density profile with $A =
0$ and $BL \simeq 7.6 \cdot 10^{-14} \, {\rm eV}$ (corresponding to
$\rho = 1 \, {\rm g/cm^3}$ and $L = 12742 \, {\rm km}$ if $A =
\frac{BL}{2}$). Parameter values:
$h = 0$, $\theta_1 = 45^\circ$, $\theta_2 = 5^\circ$, $\theta_3 = 45^\circ$,
$\Delta m^2 = 0$, and $\Delta M^2 = 3.2 \cdot 10^{-3} \, {\rm eV}^2$.}
\label{fig7}
\end{figure}

Then, in Figs.~\ref{fig6} and \ref{fig7}, we discuss linear matter
density profiles.

The parameter $B$ used in Fig.~\ref{fig6} is larger
than that in Fig.~\ref{fig7}. The smaller $B$ is the better the
agreement between the numerical evolution operator method and the
semi-classical approximation method becomes. However, linear matter
density profiles only have theoretical interest, since they are not to
be found in Nature at least at large distance scales. They could,
however, be used on shorter distances scales though, {\it e.g.}, LBL
experiments like K2K, MINOS, and CERN-LNGS \cite{LBL}, where the neutrinos
traverse the Earth mantle with cord-like paths.

\begin{figure}
\begin{center}
\includegraphics[height=8cm,angle=-90]{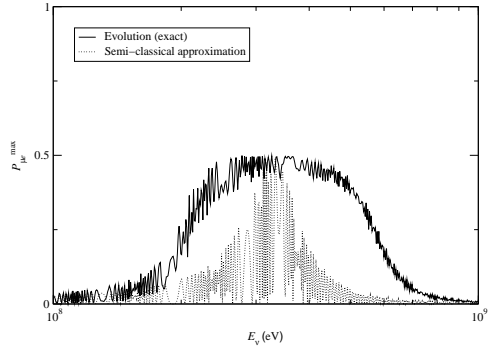}
\end{center}
\caption{The transition probability $P_{\mu e}$ as a function of the
neutrino energy $E_\nu$ for the exponentially decreasing matter density
profile of the Sun with $\rho_\odot(r) = \rho_\odot(0) e^{-r/r_0}$, where
$\rho_\odot(0) = 200 \, {\rm g/cm^3}$, $r_0 = R_\odot/10.54 \simeq
66000 \, {\rm km}$, and $R_\odot \simeq 6.96 \cdot 10^8 \, {\rm
m}$ \protect\cite{bahc89}. Parameter values: $\theta_1 =
45^\circ$, $\theta_2 = 5^\circ$, $\theta_3 = 45^\circ$, $\Delta m^2 =
0$, and $\Delta M^2 = 3.2 \cdot 10^{-3} \, {\rm eV}^2$.}
\label{fig8}
\end{figure}

Finally, in Fig.~\ref{fig8}, we investigated the Sun's matter density
profile, which is an exponentially decreasing matter density profile.
The semi-classical approximation method does not work as well
in this case as for step function, constant, and linear matter density
profiles, since this matter density profile is varying too
quickly.\footnote{A remark about the neutrino energy interval shown in
Fig.~\ref{fig8} is in place. Solar neutrinos have energies of the
order of 0.1 MeV - 10 MeV, {\it i.e.}, energies much smaller than what
is shown in Fig.~\ref{fig8}. However, the resonance energy due to the
large mass squared difference $\Delta M^2 = 3.2 \cdot 10^{-3} \, {\rm
eV}^2$ is given in Fig.~\ref{fig8}. Thus, the neutrino energy interval
shown in Fig.~\ref{fig8} is only of theoretical interest and the
figure shows how well the semi-classical approximation works in the
region where the transition probability $P_{\mu e}$ changes most.} 

In conclusion, the semi-classical approximation will be a good
approximation for some types of matter density profiles. In certain cases,
of slowly varying matter density, it is
even an excellent approximation. The major advantage of the
semi-classical approximation method as compared to the exact numerical
evolution operator method is that we only need to calculate one single
evolution operator for one single average matter density, the average
matter density parameter of the considered matter density profile $\bar{A}(L)$,
{\it i.e.}, we can make the replacement
$$
U_f(L) = \underbrace{\prod_{i = 1}^n U_f(L_i,A_i)}_{n \; {\rm operators}}
\quad \to \quad U_f(L) = \underbrace{\bar{\phi} e^{-i L \bar{T}}}_{\rm
one \; operator}.
$$

\acknowledgments

We would like to thank Martin Freund for useful comments.
This work was supported by the Swedish Foundation for International
Cooperation in Research and Higher Education (STINT) [T.O.], the
Wenner-Gren Foundations [T.O.], the
``Sonderforschungsbereich 375 f{\"u}r Astro-Teilchenphysik der
Deutschen Forschungsgemeinschaft'' [T.O.], and the Swedish Natural
Science Research Council (NFR), Contract No. F 650-19981428/2000 [H.S.].

\end{document}